\documentstyle[aps,pre,epsfig,graphicx,twocolumn]{revtex}

 \newcommand {\be} {\begin{equation}}
\newcommand {\bea} {\begin{eqnarray} \nonumber }
\newcommand {\ee} {\end{equation}}
\newcommand {\eea} {\end{eqnarray}}

\def\(({\left(}
\def\)){\right)}
\def\[[{\left[}
\def\]]{\right]}
\def\ep0{{\epsilon_0}}

\def\n0{{n_0}}

\def\vn0{{\vec {n_0}}}

\def\vm0{{\vec {m_0}}}

\begin{document}
\twocolumn[
\hsize\textwidth\columnwidth\hsize\csname@twocolumnfalse\endcsname

\title{Lattice Glass Models}
\author{Giulio Biroli$^{*}$ and Marc M{\'e}zard$^{\dagger}$}
\address{$^{*}$Center for Material Theory,
Department of Physics and Astronomy, Rutgers University, Piscataway, NJ
08854 USA\\
$^{\dagger}$Laboratoire de Physique Th{\'e}orique et Mod{\`e}les Statistiques,
Universit{\'e} Paris Sud, Bat. 100, 91405 Orsay {\sc cedex}, France }
\maketitle
\begin{abstract}
Motivated by the concept of geometrical frustration, we introduce a class
of statistical mechanics lattice models for the glass transition. 
Monte Carlo simulations in three dimensions show that 
they display a dynamical glass transition which is very similar to that 
observed in other off-lattice systems and which does not
depend on a specific dynamical rule. A mean-field study shows
the existence of a discontinuous glass transition,
in agreement with the numerical observations.
\end{abstract}
\hphantom{aaa}
]
Understanding the glass transition, and the glass phase, is one
of the present major challenges in condensed matter physics.
The experimental glass transition is related to a dramatic dynamical
slowing down in which the structural relaxation time
changes of 14 orders
 of magnitudes in a relatively small window of temperatures.
As a matter of fact, the glass transition temperature $T_g$ is 
empirically defined 
as that where the structural relaxation time becomes of the order of an
hour. In the last fifty years there have been many efforts to try to understand
whether this phenomenon is just a dramatic crossover,
it is related to a pure dynamical transition,
or it is the signature of a true thermodynamic glass
transition, called the ``ideal glass transition'', which would take place at
temperature $T_K$ below $T_g$ but which is kinetically 
avoided \cite{LibrieRiviste}. 
The possibility of this last scenario in fragile glasses
is supported by the closeness of the Vogel-Fulcher temperature
(where the extrapolated relaxation time has a divergence) and of the
Kauzmann temperature (where the extrapolated excess entropy of the supercooled
liquid vanishes)
\cite{RiAn}, and has been widely explored theoretically
\cite{LibrieRiviste,Refs,KobAndersen,FredericksonAndersen,KiThWo,MMreview}. 
The analogy with  mean field discontinuous spin 
glasses \cite{KiThWo} has given recently a new boost
to this line of research\cite{MMreview}.
These models can be solved analytically and display
striking similarities with the experimental fragile glasses.
However, they predict (because of their mean field character)
a dynamic freezing transition at a temperature $T_c>T_g$, 
equal to the mode coupling transition temperature, where real systems
still have a finite relaxation time. This difference is generally
attributed to the existence in finite dimensions of activated processes
which would transform the 'mean field' dynamical transition into a crossover
in such a way that the relaxation time increase very fast
in the temperature region $T_{K}<T<T_{c}$ and eventually diverges at $T_{K}$.

In this paper we introduce new lattice models of glasses. 
We study them numerically in three dimensions, and show
(1) that they display a dynamic glass transition similar to the one 
seen in the simulation of glass formers like binary Lennard-Jones
\cite{KobReview}, (2) that their mean field solution, obtained using the
Bethe approximation, predicts the same physical scenario found for 
mean field discontinuous spin glasses (in particular a dynamic freezing 
transition preceding the ``ideal glass transition''). 

Our lattice glass models are defined as follows. 
On each node of the lattice (e.g. a cubic lattice), there can be $0$ or $1$
particle, but these occupations are restricted by  
a hard 'density-constraint': a particle cannot have more than $\ell$ among
its $6$ neighboring sites occupied. 
One possible interpretation of this model 
is  a coarse grained
version of a usual (off-lattice) hard sphere system taking into account the 
effect of geometric frustration \cite{Nelson}. One site of our lattice is
then characterizing the density of spheres in a local cell
of space, of volume of the order of an icosahedron
built with the original spheres. 
The presence of a 'particle' in our model means that
the local arrangement of spheres in this cell is very dense,
as happens when icosahedral order sets in. The absence of the 'particle'
corresponds to a less dense arrangement of spheres. Geometric frustration
means that it is not possible to fit together the high density icosahedral
structures to fill the space: 
this is taken into account by the 'density-constraint'.

Notice the important difference with the
Kob-Andersen model \cite{KobAndersen,Kurchanetal1}.
In that case the jamming is forced by
a dynamical rule: a particle which violates the density constraint
is  blocked. 
In our case, the model is defined thermodynamically: configurations
violating the density constraint are forbidden.
The thermodynamics definition has two advantages
which were absent in previous lattice 
models\cite{KobAndersen,FredericksonAndersen} :
1) The existence of a dynamical phase transition,
and the value of the density at which it takes place, do not 
depend on the type of {\it local} dynamics which is used, e.g. whether particles 
just hop on the lattice, or whether they are exchanged 
grand-canonically with a reservoir. 2) One can perform
some analytic studies of the thermodynamics,
 relate them to the dynamical observations, and address key issues 
like  the possible existence of an ideal glass transition. 

We have run some numerical simulations of the lattice glass models
using two algorithms. 
The first one (CA) is a simple  Monte-Carlo simulation at fixed density
in the canonical ensemble,
where a randomly chosen particle can hop to a neighboring site 
if the density-constraint is satisfied.
In order to find an acceptable initial configuration, 
we prepare the system through an annealing procedure, in
which the constraint is soft: a particle with $r$ neighbors has an 
energy $E=(r-\ell)\theta(r-\ell)$ ($\theta$ is Heaviside's step function), 
and we simulate the system with Metropolis algorithm
at decreasing values of the temperature $T$. When
a zero energy configuration is found, we turn to the hard density-constraint case
(i.e. $T=0$) and start our canonical run. At equilibrium, the
chemical potential is measured from $\mu=\ln(\rho/p)$, where $p$ is the 
fraction of sites in which it is possible to add a new particle
 and $\rho $ is the density.
The second type of  simulation (GCA)
uses the grand canonical ensemble. We have a reservoir
with chemical potential $\mu$ which is coupled to each lattice site
and can create or destroy particles.

\begin{figure}[bt]
\centerline{    \epsfysize=6  cm
       \epsffile{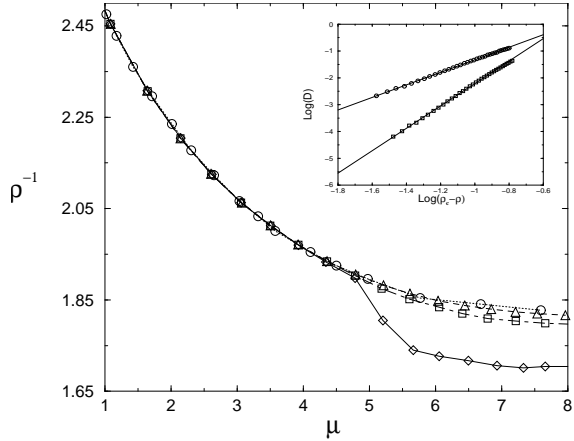}}
\caption{The inverse of  the density is plotted versus the chemical potential 
of the A particles for the mixture $m_{13}$ 
on a 3-d cubic lattice of size $15^{3}$. 
Circles are obtained from the canonical simulation CA  
with $10^3$ Monte Carlo Steps per Particle (MCSP).
Triangles, squares 
 and diamonds are obtained from the GCA 
with, respectively, $10^3,3.10^3,10^5$ MCSP at a fixed increasing rate of 
the chemical potential between $\mu =0$ and $\mu =8$. In the last case, 
the transition to the crystalline state is clearly visible.
{\bf Inset}: The diffusion coefficient for the two types of particles,
in the CA simulation of the same mixture, is plotted in log-log scale versus
$\rho_c-\rho$, showing the dynamical transition at $\rho=\rho_c$.
\label{fig_rhomu}}
\end{figure}

As expected, the GCA simulation reaches the equilibrium faster. The obtained 
results look qualitatively very similar to the ones on Lennard-Jones systems
\cite{KobReview}.
For a one component fluid,
 we find for all $\ell\ge 1$ that upon increasing $\mu$ the 
system has a first order phase
transition  towards a crystal, identified by a discontinuity of the
density, and the presence of Bragg peaks in the diffraction pattern.
In order to study the glass transition, we have considered  lattice
glass binary mixtures, for which the tendency to crystallization is reduced.
In these mixtures, one obtains a reproducible 
``supercooled liquid'' which exhibits, when the density increases,
 a dynamical glass transition. We present  the results obtained
for a mixture containing $30 \%$ of particles $A$ with density-constraint
$\ell_A=1$, and $70 \%$  of particles $B$ with density-constraint
$\ell_B=3$ (called $m_{13}$ mixture in the following).
Fig.\ref{fig_rhomu} shows the density as a function of the chemical 
potential,
measured in the two algorithms. One sees a clear saturation
which takes place at a density $\rho_c \simeq 0.565$: none of these two local
algorithms can reach a density higher than $\rho_c$. The Inset
of Fig.\ref{fig_rhomu} shows the diffusion coefficient $D_A$ (resp. $D_B$)
of the particles measured in the
CA simulation. Their decrease with increasing $\rho$
 is well fitted by a function
vanishing at $\rho_c$ with a power law:
\be
D_{A,B}(\rho)\simeq C_{A,B} (\rho_{c}-\rho)^{\alpha_{A,B}} \theta(\rho_{c}-\rho)
\ee
with exponents of order $\alpha_{A}\sim 4.2$ and $\alpha _{B}\sim 2.3$.

This vanishing is generally taken as the numerical signature of the dynamical 
phase transition \cite{careful,susce}.
When the chemical potential is quenched above the critical value 
(corresponding to the dynamical transition) the systems remains 
out of equilibrium and has an aging behavior 
which we have seen measuring the two time correlations
of occupation numbers of a site.
On the contrary, if we increase
the chemical potential very slowly, we can observe the transition to a
crystalline phase even in this case of mixtures as shown in Fig. \ref{fig_rhomu}.
	It is quite possible that the $m_{13}$ mixture will eventually phase separate
on very large times, but we have not seen any such effect on the time scales
of our simulations.
 Whether there exist
mixtures where the true equilibrium state at large $\mu$ is a glass,
with a thermodynamic phase transition from the liquid to this glass,
is under current study \cite{BirMezPar}. Let us finally remark
that the existence of a metastable glassy phase does not depend
on the type of {\it local} dynamics we use (GCA or CA), only the 
time to nucleate the crystal does (contrary to the 
Kob-Andersen model).  

We now present a mean field theory of the finite dimensional
lattice glass models. We focus on the Bethe
lattice version of these models
\cite{BetheSG}, 
whose underlying lattice structure
is a random graph with fixed connectivity: every
vertex has exactly  $k+1$ neighbors, but the graph is otherwise random 
(to study the three dimensional case one takes $k=5$).
Locally (on finite lenghtscales), such a graph has the structure
of a Cayley tree with a fixed branching ratio, but it  
 also has loops of typical size $\ln N$. The presence of these loops
is crucial to induce the geometric frustration,
but the local tree structure allows for an analytic solution of the
model. 
For the sake of clarity, we shall describe this solution in the simple 
case where there is only one type of particle,
with a density-constraint given by the integer $\ell$; we have
extended these computations to mixtures \cite{Long} and we shall present 
the results for the $m_{13}$ mixture.

Since the underlying
lattice structure is locally  tree-like, one can write iterative 
equations on the local probability measure.
More precisely, let us analyze one branch of the tree
ending on site $i$. We denote by   $j \in \{ 1,...,k\} $
all the neighbors of $i$. We call
 $Z_0^{(i)}$ the partition function of this branch, restricted
to configurations where site $i$ is empty, and $Z_{1,p}^{(i)}$ 
the partition function restricted
to configurations where site $i$ is occupied, and has $p$ neighbors occupied.
Defining $S_{q}^{(i)}\equiv \sum_{p=0}^q Z_{1,p}^{(i)}$, we find
the recursion relations:
\bea
Z_0^{(i)}&=&\prod_{j=1}^k \(( Z_0^{(j)}+S_{\ell}^{(j)} \)) \\
Z_{1,p}^{(i)}&=& e^\mu \sum_{1\le j_1<...<j_p\le k}S_{\ell-1}^{(j_1)}...
S_{\ell-1}^{(j_p)} \prod_{m \notin \{j_1,...,j_p\}} Z_0^{(m)} \ .
\label{iterZ}
\eea
It is convenient to introduce on any site $m$ the local fields
$h_m=\ln\((S_{\ell}^{(m)}/Z_0^{(m)}\))$ and
$a_m=\ln\((S_{\ell-1}^{(m)}/Z_0^{(m)}\))-h_m$, in terms of which the 
iteration reads:
\be\label{eqh}
e^{h_{i}}=e^\mu \ \((\prod_{j=1}^k {1 \over 1+e^{h_j}}\)) \ 
\sigma_\ell
\ \ , \ \ 
e^{a_{i}}=\frac{\sigma_{\ell-1}}{\sigma_{\ell}} \
\ee
where 
\be
\sigma_{q}=\sum_{p=0}^q
\[[
\sum_{1\le j_1<...<j_p\le k}e^{h_{j_1}+a_{j_1}+...+h_{j_p}+a_{j_p}} \]]
\ee
and the sum over $j_1<...<j_p$ is defined to take the value $1$ when $p$ is
 zero.

 From these fields, one can obtain \cite{Long} the grand canonical 
potential $A=-\log(Z)$ as a sum of sites and links contribution 
\cite{Katsura,BetheSG}: $A=-k \sum_{i} A_{i}^{(1)}
+\sum_{<i,i'>}A_{i,i'}^{(2)}$. The contribution 
from the bond $<i,i'>$ is given by the local fields obtained 
from the two branches
arriving on $i$ (in absence of $i'$) and on  $i'$ (in absence of $i$) as:
\begin{equation}\label{bondcontribution}
\exp\((-A_{i,i'}^{(2)}\))=1+e^{h_i}+e^{h_{i'}}+e^{h_i+a_i+h_{i'}+a_{i'}}
\end{equation}
 The contribution  from site $i$ reads
\begin{equation}\label{sitecontribution}
\exp\((-A_{i}^{(1)}\))=1+e^{H_i}
\end{equation}
where $H_i$ is the total field on site $i$,
given by formula (\ref {eqh}) where $k$ is changed to $k+1$.
The shift in grand potential when merging the $k$ branches ending at points $j=1,...,k$ onto the node $i$ is given by
\begin{equation}
\exp\((-A_{i}'\))= 1+e^{h_i} \ .
\label{aiter}
\end{equation}
\begin{figure}[bt]
\centerline{    \epsfysize=5  cm
       \epsffile{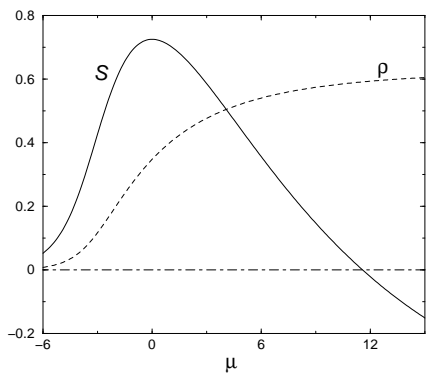}}
\caption{The entropy per site (continuous line) and the total density (dashed
 line) predicted by the liquid solution 
are plotted as a function of the chemical potential of the $A$ 
particles for the mixture $m_{13}$ on a $k=5$ Bethe lattice.
\label{fig:entro}}
\end{figure}

Starting from  the previous eqs., 
we find  at low density a liquid phase 
characterized by a homogeneous solution $h_{i}=h,a_{i}=a$,
where  $h,a$ verify two simple self-consistent equations. 
Given this solution  one can easily obtain 
all the thermodynamic quantities, and in particular the 
density $\rho $ and the entropy per lattice site $S=A-\rho \mu$. 
For every  $k\ge 2$
and $\ell \leq k$, 
(and for generic mixtures), the entropy becomes negative 
when the chemical potential $\mu $ becomes larger than a certain 
critical value $\mu_{s=0}$. Therefore  a thermodynamic
phase transition  takes place at a chemical potential
 $\mu\le\mu _{s=0}$, as shown in  Fig. \ref{fig:entro}.

To gain some further insight on the thermodynamic transition we study the stability
of the liquid phase, i.e. we analyze all the generalized susceptibilities:
\begin{equation}\label{susceptibilities}
\Xi_{p}=\frac{1}{N}\sum_{i,j}\overline{<n_{i}n_{j}>_{c}^{p}}
=\sum_{d}k^{d}(G_{d})^{p}
\end{equation}
where $n_i \in \{0,1\} $ is the occupation number of site $i$ and
$<n_{i}n_{j}>_{c}=G_d$ denotes the connected correlation function between
points $i$ and $j$ at a distance $d$. 
Since there is 
only one finite path (with probability one) 
connecting two points at a finite distance $d$ , 
the computation of $G_{d}$ (and therefore the stability analysis) can be reduced
to a one dimensional problem which can be solved by transfer matrix technique
\cite{Long}.
The divergence of $\Xi_{1}$ signals an 
instability toward a cyclic solution of the iteration eqs. \cite{Long}, 
which corresponds to the crystal as found for $l=0$ by Runnels \cite{Runnels}.
Here, we focus on the next instability related to 
the divergence of the  ``glass susceptibility'' $\Xi_{2}$.
Depending on the value of $k,l$ (and the type of mixture), we can encounter 
two types of situations. The first one is when 
the  glass susceptibility remains always finite 
(this is found for $k=1,2,3$ and every $l>0$), the glass transition is
then  discontinuous.
Instead, when the susceptibility $\Xi_{2}$ diverges 
at $\mu=\mu_2$, either there is a continuous glass phase transition 
 at $\mu=\mu_2$, (this is found for
$\ell=0$, which is nothing but the vertex covering model studied
in \cite{VerCov}), or a discontinuous phase transition 
(without any divergence of $\Xi_2$)  takes place at 
$\mu_c<\mu_2$. This happens for sure
whenever $\mu _{s=0}<\mu_2$ (this is found e.g. in the $k=5,l=3$ case). 
The discontinuous transition case,
on which we focus in the following, is the one relevant for glasses, 
contrary to the continuous transition which is typical of 
spin glasses \cite{MMreview}. 

The high-density glassy phase can be analyzed 
in the Bethe approximation, taking into account
the existence of many different local minima of the TAP
free energy (called pure states in the following, see \cite{Beyond}).
In this case the fields
$h_{i}$ and $a_{i}$  fluctuate not only from site to site 
but also from pure state to pure state. One  defines for each site a probability 
distribution $R_{i}(h,a)$ that the fields $h_{i},a_{i}$ equal
$h,a$ for a randomly chosen pure state. 
In our case we have verified that $R_{i}$ does 
not fluctuate from site to site, and the analysis of 
the high-density glassy phase reduces to obtaining a single function $R(h,a)$.
Using the cavity or the replica method \cite{Long}  we find that this function 
satisfies the self-consistent equation:
\begin{equation}\label{onersbeq}
\frac{{\cal N}R(h,a)}{(1+e^h)^{-m}}=\int \prod _{j=1}^{k}\[[dh_{j}da_{j}R(h_{j},a_{j})\]]\delta \left(h-h_{i} \right) \delta \left(a-a_{i} \right) 
\end{equation}
where ${\cal N}$ is a normalization constant, $h_{i}$, $a_{i}$ are the local fields 
on site $i$
obtained when merging $k$ branches which carry the 
fields $\{ h_{j},a_{j} \}$, see eq. (\ref{eqh}), and $m$ is
 a Lagrange multiplier which fixes the value of the free energy density 
of the pure states giving rise to $R(h,a)$ \cite{RemiPRL}.
We have solved this eq. numerically using the algorithm 
of \cite{BetheSG}.
For  some choices of $k$ and $\ell$, we find a scenario identical to the one
of  discontinuous spin glasses,
with two transitions:
(1)  a  dynamical transition at $\mu _{d},\rho _{d}$,
(2)  an equilibrium glass transition, due to an entropy crisis {\`a} la Kauzmann,
at a certain density $\rho _{eq}>\rho _{d}$ and chemical potential $\mu _{eq}>\mu _{d}$. 
Precisely, when we increase $\mu$ starting from the liquid phase (where
$R_{liq}(h,a)=\delta(h_{liq}-h)\delta(a_{liq}-a)$),
we first encounter at 
 $\mu=\mu_{d}$ a non trivial solution $R_{glass}^{m=1}(h,a)$ 
of (\ref{onersbeq}) which
appears discontinuously, signaling the existence of many pure states.
 The static (equilibrium) transition appears 
at a higher chemical potential when these new solutions dominate the
thermodynamics\cite{RemiPRL,MMreview}.
For the $m_{13}$ mixture on a $k=5$ lattice, we find  $\rho _{d}\simeq 0.58$,
which  is surprisingly close to the 3D value. 
The discontinuous
character of the transition is assured by the fact that the
difference $q_{1}-q_{0}$ is finite and positive at the transition, where
the overlaps $q_{0}$ and $q_{1}$ are defined as usual 
\cite{Beyond}: $q_{0}=(1/N)\sum_{i }
<n_{i}>_{\alpha }<n_{i}>_{\beta }/N$ 
and $q_{1}=(1/N)\sum_{i }<n_{i}>_{\alpha }^{2}/N$ 
and the indices $\alpha ,\beta $ 
denote two pure states randomly chosen according to their Boltzmann weights.
Moreover we have checked, by running a CA simulation on
the Bethe lattice \cite{Long}, that the dynamical transition
takes place at the density where the diffusion coefficients 
vanish.
A detailed study of the thermodynamics
of the lattice glass models, and of their equilibrium transition,
will be presented elsewhere \cite{BirMezPar}.


In this letter we have introduced
a new class of lattice models of glasses. Their 3D numerical simulations 
display a phenomenology very similar to the one 
of the Lennard-Jones systems, whereas their solution 
in the Bethe approximation exhibits a discontinuous 
glass transition. These
models allow to bridge the gap between the 
phenomenology of fragile glasses and their disordered spin glass analogs.
They should allow to study  key issues like the
existence or not of a true thermodynamic phase transition,
and the finite dimensional counterpart of the mean field dynamical transition,
 because
(1) they are among the simplest systems to exhibit a glass transition 
in 3D, (2) this transition is not linked to a specific dynamical rule,
but it is present in any local dynamics, (3) contrary to off-lattice models, 
they can be solved in the Bethe approximation (i.e. in the limit of infinite
dimension), (4) within such an approximation scheme 
they present a discontinuous spin glass transition.

Acknowledgments: We warmly thank G.Parisi for many useful remarks and 
comments, and S. Franz and F. Ricci for interesting discussions.


\begin{thebibliography}{99}
\bibitem{careful} As in any solid, 
 particles always eventually diffuse because of vacancies, 
but this diffusion is not seen on the time
scales of numerical simulations.
\bibitem{susce} We have checked that the spin glass susceptibility stays 
finite at $\rho _{c}$. This rules out the possibility that the dynamical 
transition is accompanied by a spin glass like 
thermodynamical phase transition.
\end{thebibliography}
\end{document}